\documentclass[referee,usenatbib,usegraphicx]{biom}

\usepackage{amsmath,amssymb,amsfonts,color,soul}
\usepackage{booktabs}
\usepackage{caption}
\usepackage{subfig}
\usepackage{enumerate}
\usepackage{rotating,graphicx,multicol}
\usepackage[ruled]{algorithm2e}

\DeclareMathOperator{\dev}{d\!}
\def\bSig\mathbf{\Sigma}
\newcommand{\joinR}{\hspace{-.1em}}
\newcommand{\RomanI}{I}
\newcommand{\RomanII}{\mbox{\RomanI\joinR\RomanI}}
\newcommand{\RomanIII}{\mbox{\RomanI\joinR\RomanII}}
\newcommand{\RomanIV}{\mbox{\RomanI\joinR\RomanV}}
\newcommand{\RomanV}{V}
\newcommand{\RomanVI}{\mbox{\RomanV\joinR\RomanI}}
\newcommand{\RomanVII}{\mbox{\RomanV\joinR\RomanII}}

\newcommand{\tT}{{\mathcal{T}}}
\newcommand{\mT}{{\mathcal{T}}}
\newcommand{\T}{\mathcal{T}}
\newcommand{\AUC}{\text{AUC}}
\newcommand{\CON}{\text{CON}}
\newcommand{\TPR}{\text{TPR}}
\newcommand{\FPR}{\text{FPR}}
\newcommand{\ROC}{\text{ROC}}
\newcommand{\ICON}{\text{ICON}}

\newcommand{\hP}{\widehat{P}}

\newcommand{\bZ}{{\bm Z}}
\newcommand{\bX}{{\bm X}}
\newcommand{\bx}{{\bm x}}
\newcommand{\bz}{{\bm z}}
\newcommand{\bbeta}{{\bm \beta}}

\theoremstyle{plain}
\newtheorem{thm}{Theorem} 
\newtheorem{proposition}[thm]{Proposition}

\title[ROC-Guided Survival Trees and Ensembles]{ROC-Guided Survival Trees and Ensembles}

\author
{Yifei Sun$^1$\email{ys3072@cumc.columbia.edu},
 Sy Han Chiou$^2$, and Mei-Cheng Wang$^3$ \\
$^1$ Department of Biostatistics, Columbia Mailman School of Public Health,\\ New York, New York 10032, U.S.A.\\
$^2$ Department of Mathematical Sciences, University of Texas at Dallas,\\ Richardson, Texas 75080, U.S.A.\\
$^3$ Department of Biotatistics, Johns Hopkins Bloomberg School of Public Health, \\
Baltimore, Maryland 21205, U.S.A.
}

\begin{document}

\pagerange{\pageref{firstpage}--\pageref{lastpage}}

\doi{}

\label{firstpage}

\begin{abstract}
Tree-based methods are popular nonparametric tools in studying time-to-event outcomes. In this article, we introduce a novel framework for survival trees and ensembles, where the trees partition the dynamic survivor population and can handle time-dependent covariates. Using the idea of randomized tests, we develop generalized time-dependent Receiver Operating Characteristic (ROC) curves for evaluating the performance of survival trees. The tree-building algorithm is guided by decision-theoretic criteria based on ROC, targeting specifically for prediction accuracy. To address the instability issue of a single tree, we propose a novel ensemble procedure based on averaging martingale estimating equations, which is different from existing methods that average the predicted survival or cumulative hazard functions from individual trees. Extensive simulation studies are conducted to examine the performance of the proposed methods. We apply the methods to a study on AIDS for illustration. 
\end{abstract}


\maketitle

\section{Introduction}
Tree-based methods are popular alternatives to semiparametric and parametric methods. The basic idea of tree-based methods is to partition the covariate space into subsets (nodes) where individuals in the same node are alike regarding the outcome of interest. A single prediction is then assigned to individuals in the same node. In the setting of classification and regression trees (CART) \citep{breiman1984classification}, this can be achieved by greedy splitting algorithms that minimize a measure of node impurity and the sum of squared deviations from the node mean, respectively. To improve predictive accuracy and reduce overfitting, one can apply a cost-complexity pruning algorithm to determine the size of the tree. In practice, however, a small perturbation in the data may result in a substantial change in the structure of a fitted tree. Ensemble methods \citep{breiman1996bagging,breiman2001random} are ideal solutions to the instability problem and outperform a single tree in many applications. 

With the increasing focus on personalized risk prediction, survival trees for time-to-event data have received much attention. A survival tree partitions the predictor space into a set of terminal nodes and reports the Kaplan-Meier estimate of the survival function in each node. In the literature, there has been a steady stream of works proposing new splitting rules to build survival trees. \cite{gordon1985tree} first adopted the idea of CART and defined the node impurity to be the minimum Wasserstein distance between the Kaplan-Meier curves of the current node and a pure node. Alternatively, splitting rules that maximize the between node heterogeneity were commonly used in the literature: for example, \citet{ciampi1986stratification}, \citet{segal1988regression}, \citet{leblanc1993survival} suggested  
selecting a split that yields the largest log-rank statistic; \cite{moradian2017l_1} used the integrated absolute difference between survival functions to measure the dissimilarity between child nodes. Moreover, researchers also considered likelihood-based splitting criteria, where the split is selected to maximize the sum of log-likelihoods from the two child nodes \citep{leblanc1992relative}. Recently, the squared error loss in regression trees was extended to censored data \citep{molinaro2004tree, steingrimsson2016doubly, steingrimsson2018censoring}. Other splitting criteria include a weighted sum of impurity of the censoring indicator and the squared error loss of the observed event time \citep{zhang1995splitting}, and the Harrell's C-statistic \citep{schmid2016use}. Readers are referred to \cite{bou2011review} for a comprehensive review of survival trees. 

To address the instability issue of a single survival tree, researchers have developed various ensemble methods for time-to-event data. The basic algorithm operates by combining a large number of survival trees constructed from resampling the training data. For example, \cite{hothorn2004bagging} proposed a general method for bagging survival trees, and the final Kaplan-Meier estimate is computed using aggregated observations from all individual trees; \cite{hothorn2006survival} proposed a random forest method to predict the log survival time; \cite{ishwaran2008random} proposed the random survival forest, where the estimates of the cumulative hazard function are averaged for the final prediction. \cite{zhu2012recursively} used extremely randomized trees and proposed an imputation procedure that recursively updates the censored observations. \cite{steingrimsson2018censoring} considered more general weighted bootstrap procedures. Theoretical properties of random survival forests have been studied in \cite{ishwaran2010consistency} and \cite{cui2017some}.

The use of time-dependent covariates offers opportunities for exploring the association between failure events and risk factors that change over time. Applying standard methods such as the Cox proportional hazards model could be challenging in the choice of covariate form and can often yield biased estimation  \citep{fisher1999time}. In contrast, tree-based methods allow the event risk to depend on the covariates flexibly.  \cite{bacchetti1995survival} incorporated time-dependent covariates by using ``pseudo-subject'', where the survival experience of one subject is viewed as experiences of multiple pseudo-subjects on non-overlapping intervals, and survival probability using the truncation product-limit estimator is reported as the node summary. The idea of pseudo-subject was employed by most of the existing works dealing with time-dependent covariates \citep{huang1998piecewise,bou2011discrete,wallace2014time,fu2017survival}.  The pseudo-subject approach may have practical limitations, because one subject could be classified into multiple nodes in a tree, leading to a loss of simple interpretation and possible ambiguous prediction. 

In this article, we propose a unified framework for survival trees and ensembles. {To incorporate time-dependent covariates, we propose a time-invariant partition scheme on the survivor population. The partition-based risk prediction function is constructed using an algorithm guided by Receiver Operating Characteristic (ROC) curves. Specifically, we define generalized time-dependent ROC curves and show that the target hazard function yields the largest area under the ROC curve. The optimality of the target hazard function motivates us to use a weighted average of the areas under the curve (AUC) on a time interval} to guide splitting and pruning. Finally, we propose a novel ensemble method that averages unbiased martingale estimating equations instead of survival predictions from individual trees.

\vspace{-1cm}
\section{Survival trees with time-dependent covariates}
\subsection{A time-invariant partition on survivor population}
Suppose $T$ is a continuous survival time and $\bZ(t)$ is a $p$-dimensional vector of possibly time-dependent covariates. Denote by {$\lambda(t\mid \bz)$} the hazard function of $T$ given $\bZ(t) = \bz$, i.e.,
\begin{align}
\label{haz}
\lambda(t\mid \bz)\dev t = P\{T\in[t,t+\dev t) \mid \bZ(t) = \bz, T\ge t \}.
\end{align}
The function $\lambda(t\mid \bz)$ characterizes the instantaneous failure risk at $t$ among survivors. At time $t$, let $\mathcal{Z}_t$ denote the covariate space of $\bZ(t)$ in the survivor population (i.e., the subpopulation satisfying $T\ge t$). To illustrate the idea, we first assume $\mathcal{Z}_t=[0,1]^p$ for $t\in (0,s]$ and $s$ is a pre-specified constant. We consider a partition on $\mathcal{Z}_t$ that divides the survivor population into $M$ groups, denoted by $\T=\{ \tau_1, \ldots, \tau_M\}$. The partition $\T$ is time-invariant and can be applied on $\mathcal{Z}_t$ for all $t\in (0, s]$. The elements of the partition are disjoint subsets of $\mathcal{Z}_t$ satisfying $\bigcup_{m = 1}^M\tau_m = \mathcal{Z}_t$ and are called terminal nodes. A subject enters a terminal node $\tau$ at $t$ if $\bZ(t)\in\tau$ and $T\ge t$. The partition $\T$ induces the following hazard model,
\begin{align}
\label{m1}
\lambda_{\T}(t\mid \bZ(t)) = \sum_{\tau\in \T}I(\bZ(t)\in\tau) \lambda(t\mid\tau),  ~ 0<t\le s,
\end{align}
where $\lambda(t\mid\tau) \dev t= P\{T\in[t,t+\dev t) \mid \bZ(t) \in\tau, T\ge t \}$ is the node-specific hazard. Define a partition function $l_{\T}$ so that $l_{\T}\{\bz\} = \tau$ if and only if $\bz\in\tau$, $\tau\in\T$. The partition-based hazard function can then be written as $\lambda_{\T}(t\mid \bz) = \lambda(t\mid l_{\T}\{ \bz\})$.

\begin{figure}
    \centering
    \includegraphics[scale = 0.8]{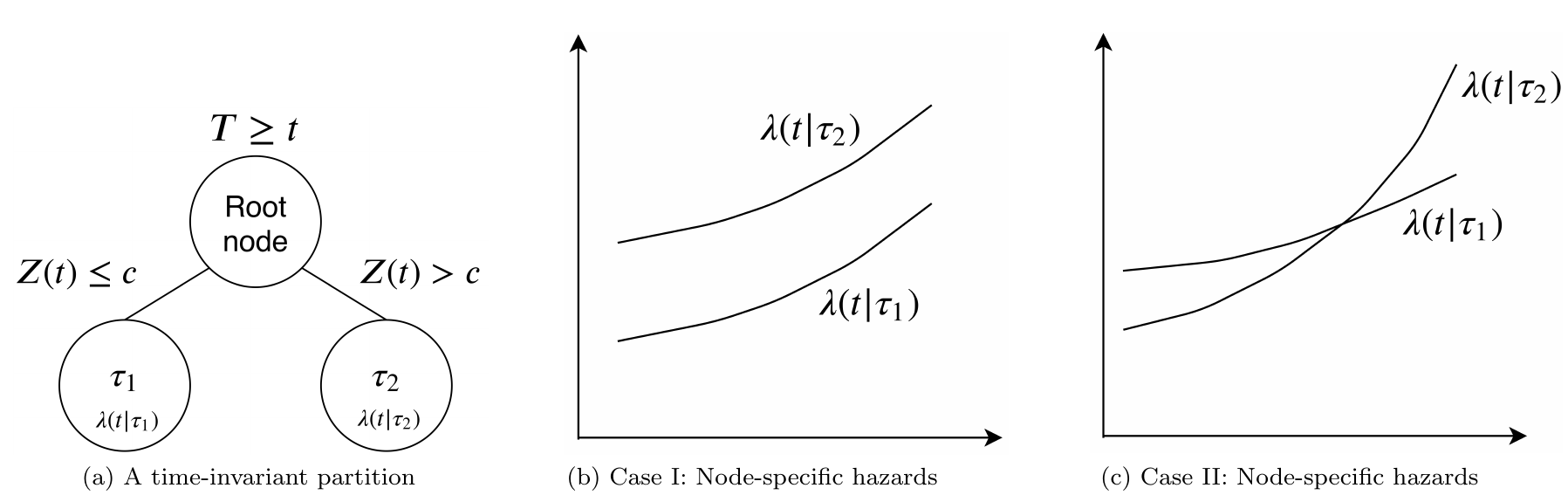}
	\caption{Illustration of the survival tree and hazard prediction.}
	\label{fig1}
\end{figure}

The time-invariant partition allows a sparse model and an easy interpretation of the decision rule. At each time $t$, the tree partitions the survivor population based on $\bZ(t)$ and predicts the instantaneous failure risk. Thus the interpretation at a fixed time point is along the same line as classification and regression trees. Since the risk within each terminal node changes with time, it is essential to look at the hazard curves of each terminal node to determine the subgroups with high failure risks. Consider an example in Figure~\ref{fig1}a where the partition $\T = \{\tau_1,\tau_2 \}$ based on a single predictor (i.e., $p=1$) divides the survivor population into two subgroups. Figures~\ref{fig1}b and~\ref{fig1}c are two possible scenarios the node-specific hazards can be defined. A larger value of the predictor is associated with a higher risk if the node-specific hazards are in Figure~\ref{fig1}b, while a larger value of the predictor is associated with a lower risk in the early period and higher risk in the later period if the node-specific hazards are in Figure~\ref{fig1}c. The partition-based hazard function $\lambda_{\T}(t\mid \bz)$ approximates the true hazard function $\lambda(t\mid \bz)$ as the partition becomes more refined.

\newcommand{\hF}{\widehat{F}}
\begin{remark}
\label{rmk1}
The above discussion focuses on the case where ${\mathcal Z}_t = [0,1]^p$ for $t\in(0,s]$. In practice, if the domain of the time-dependent covariates among survivors changes over time, a time-invariant partition may not be appropriate. In this case, one can transform $\bZ(t)$ onto $[0,1]^p$ via a one-to-one function $G_t:\mathcal{Z}_t\mapsto[0,1]^p$. Let $\bX(t) = G_t(\bZ(t))$ be the transformed covariates and $h(t\mid \bx)$ be the hazard function of $T$ given $\bX(t) = \bx$. A tree $\T'$ can be constructed using the transformed covariates $\bX(t)$. Since $\lambda(t\mid \bz) = h(t\mid G_t(\bz))$, the tree-based hazard given $\bZ(t) = \bz$ is ${\lambda}_{\T'}(t\mid G_t(\bz))$. One may use  $G_t(\bz) {=} \widehat{F}_t(\bz) \overset{\rm def}{=} (\widehat{F}_{1t}(z_1),\ldots,\widehat{F}_{pt}(z_p))$, where $\bz = (z_1,\ldots,z_p)$ and for $q = 1,\ldots, p$, $\widehat{F}_{qt}$ is the empirical cumulative distribution function of the $q$th element of $\bZ(t)$ among the at-risk subjects (i.e., $Y\ge t$). 
\end{remark}

\subsection{Estimation of tree-based risk function}

Given a partition $\T$, we now consider the estimation of $\lambda_{\T}$ with right-censored survival data. Let $Y = \min(T,C)$ be the observed survival time and $\Delta = I(T\le C)$ be the failure event indicator. We use $\bZ^H(t) = \{\bZ(u), 0\le u\le t \}$ to denote the covariate history up to $t$. The training data, denoted by $\mathcal{L}_n = \{Y_i,\Delta_i,\bZ_i^H(Y_i); i = 1,\ldots,n \}$, are independent identically distributed (i.i.d.) replicates of $\{Y,\Delta,\bZ^H(Y) \}$. To facilitate the discussion, we start from the case where there are a fixed number of terminal nodes in $\mT$ and assume independent censoring within each terminal node, that is, $P\{ T\in [t,t+\dev t) \mid \bZ(t)\in\tau, T\ge t, C\ge t \} = \lambda(t\mid\tau)\dev t$ for $\tau\in\T$. As will be discussed later, a conditionally independent censoring assumption is imposed when the number of terminal nodes is allowed to increase with $n$. {For $\tau\in\T$, define $F^{\ast}(t,\tau) = P\{Y\le t, \Delta = 1, \bZ(Y)\in\tau \}$ and $S^{\ast}(t,\tau) = P\{\bZ(t)\in\tau, Y\ge t  \}$. } Then we have
$\lambda(t\mid  \tau) ={f^{\ast}(t,\tau)}/{S^{\ast}(t,\tau)}$,
where $f^{\ast}(t,\tau) = \dev F^{\ast}(t,\tau)/\dev t$. Define the observed counting process $N(t) = \Delta I(Y\le t)$.  We estimate $f^{\ast}(t,\tau)$ by the following kernel type estimator,
\begin{align*}
\widehat f^{\ast}(t,\tau) = \int_{0}^{s} K_h(t-u) \dev \widehat{F}^{\ast}(u,\tau), ~~t \in [h,s-h],
\end{align*}
where $\dev\widehat{F}^{\ast}(u,\tau) = \sum_{i=1}^n I({\bZ}_i(Y_i)\in\tau)\dev N_i(u)/n$, $K_h(\cdot) = K(\cdot/h)/h$, $K(\cdot)$ is a second order kernel function with a support on $[-1,1]$, and $h$ is the bandwidth parameter. To avoid biased estimation in the boundary region, one can either use the second order boundary kernel  \citep{muller1991smooth}, or set $\widehat f^{\ast}(t,\tau) = \widehat f^{\ast}(h,\tau)$ for $t\in[0,h)$ and  $\widehat f^{\ast}(t,\tau) = \widehat f^{\ast}(s-h,\tau)$ for $t\in(s-h,s]$. Given a node $\tau$, $\widehat{f}^{\ast}(t,\tau)$ consistently estimates ${f}^{\ast}(t,\tau)$ as $n\rightarrow\infty$, $h\rightarrow 0$ and $nh\rightarrow \infty$. Note that $S^{\ast}(t,\tau)$ can be straightforwardly estimated by $\widehat{S}^{\ast}(t,\tau) = \sum_{i=1}^n I(\bZ_i(t)\in\tau, Y_i\ge t)/n$. Thus the node-specific hazard ${\lambda}(t\mid\tau)$ can be estimated by 
\begin{align*}
\widehat{\lambda}(t\mid\tau){=} {\int_0^{\infty} K_h(t-u) \dev \widehat{F}^{\ast}(u,\tau)}\big/{\widehat{S}^{\ast}(t,\tau)},
\end{align*}
and ${\lambda}_{\T}(t\mid \bz)$ can be estimated by $\widehat\lambda_{\T}(t\mid \bz)=\widehat{\lambda}(t\mid l_{\T}\{\bz\})$.

In practice, the partition is usually constructed from data dependent algorithms. To study the large-sample property of the predicted hazard, we use $\T_n$ to denote a partition that can depend on the training data and define  $\widehat{\lambda}_{\T_n}(t\mid \bz) = \widehat{\lambda}(t\mid l_{\T_n}\{\bz\}) $. Given a new observation $\bZ_0(t)$ that is independent of the training data, we predict the hazard to be $\widehat{\lambda}_{\T_n}(t\mid \bZ_0(t))$.  We assume that the following conditions hold:
\begin{description}
	\item[(A1)] The censoring time $C$ satisfies $P(t\le T<t+\dev t\mid\bZ(t)=\bz,T\ge t, C\ge t) = \lambda(t\mid \bz)\dev t$. 
	\item[(A2)] The process $\bZ(t)$ is left-continuous and has right-hand limit. At time $t$, $\bZ(t)$ is distributed to a bounded density on $[0,1]^p$. There exists a constant $c_1$ such that $f_{\bZ(t)\mid Y\ge t}(\bz)P(Y\ge t)>c_1$ for $\bz\in[0,1]^p,t\in[0,s]$, where $f_{\bZ(t)\mid Y\ge t}(\bz)$ is the density of $\bZ(t)$ given $Y\ge t$. 
	\item[(A3)] The function $f^{\ast}(t\mid\bz)=\lim_{\delta_t\rightarrow 0^+}P\{ T\in[t,t+\delta_t),\Delta=1\mid \bZ(Y)=\bz \}/\delta_t$ is second order differentiable with respect to $t$, and $\sup_{t\in[0,s],\bz\in[0,1]^p}|\partial^2f^{\ast}(t\mid\bz) /\partial t^2|<c_2$ for some constant $c_2$. 
	\item[(A4)] The number of terminal nodes in $\T_n$ grows at the rate of $o({nh}/{\log n})$. For any $\gamma>0$, the diameters of the nodes satisfies $\mu(\bz:\text{diam}(l_{\T_n}\{\bz\})>\gamma) \rightarrow 0$ with probability 1, where $\mu$ is the Lebesgue measure.
	\item[(A5)] The bandwidth $h$ satisfies $h = n^{-\alpha}$, $0< \alpha < 1$.
\end{description}
The conditional independent censoring assumption (A1) has been commonly adopted in survival analysis. Condition (A2) requires the support of $\bZ(t)$ in the at-risk population to be time-invariant on the time interval of interest. Condition (A3) requires the target function to be smooth so that kernel smoothing can be applied. Condition (A4) allows the size of the tree to become larger as the sample size increases. An analogous condition can be found in \cite{nobel1996histogram}, where the number of terminal nodes grows in the rate of $o(n/\log n)$ for regression trees. The size of the tree is smaller in our case as our hazard estimator is based on kernel smoothing. The shrinking cell condition in (A4) is analogous to (12.9) in \cite{breiman1984classification}. Condition (A5) implies $h\rightarrow 0$ and $nh/\log n\rightarrow\infty$ as $n\rightarrow\infty$. We require the diameters of the terminal nodes and the bandwidth $h$ to shrink towards zero so that the local estimate can approximate target function as $n\rightarrow\infty$. Similar to the existing works on tree-based estimators, the convergence result does not account for any specific splitting rule or pruning procedure. Theorem~\ref{thm} is developed to justify the hazard prediction under a set of regularity conditions. The proof is given in the Supporting Information. 

\vspace{-0.2cm}
\begin{theorem}
	\label{thm}
	Under conditions (A1)-(A5), for $0<t\le s$ and any $\epsilon>0$, as $n\rightarrow\infty$, we have $P\left\{ \left|\widehat{\lambda}_{\T_n}(t\mid \bZ_0(t))-\lambda(t\mid \bZ_0(t))\right|>\epsilon\mid \mathcal{L}_n\right\} \rightarrow 0$ with probability 1. 
\end{theorem}

Although the above discussion focuses on the hazard function, one can also predict the survival probability when $\bZ(t)$ is a vector of \emph{external} time-dependent covariates \citep{kalbfleisch2011statistical}. Examples of external time-dependent covariates include fixed covariates, covariates determined in advance for each individual, and ancillary covariates whose probability laws do not involve parameters in the failure time model. Assume the hazard at $t$ depend on $\bZ^H(t)$ only through $\bZ(t)$. The prediction of survival probability is based on the equation $P(T\ge t\mid \bZ^H(t)) = \exp\{ -\int_{0}^{t}\lambda(u\mid \bZ(u))\dev u \}$.  We predict the survival probability at $t$ for a subject with covariate path $\bZ_0^H(t) = \{\bZ_0(u),0\le u\le t \}$ to be
\begin{align}
\label{spred}
\hP(T\ge t\mid \bZ_0^H(t))=\exp\left[-\int_{0}^{t} \frac{\sum_{i=1}^{n}I(\bZ_i(u)\in l_{\T}\{\bZ_0(u)\})\dev N_i(u)}{\sum_{i=1}^{n} I(\bZ_i(u)\in l_{\T}\{\bZ_0(u)\}, Y_i\ge u)} \right].
\end{align}

\vspace{-.2in}

\section{ROC-Guided survival trees}
In this section, we propose an ROC-guided algorithm where the partition is constructed via a greedy approach that aims to maximize an ROC-related measure. {The incident/dynamic time-dependent ROC curve \citep{heagerty2005survival} is a popular tool for evaluating the prognostic accuracy of a continuous marker. Heuristically, at time $t$, the ROC curve is defined on the survivor population, where a subject is considered a case if $T=t$ and a control if $T> t$. Let $g(\cdot): \mathcal{Z}_t\mapsto\mathbb{R}$ be a scalar function  that summarizes information from $\bZ(t)$, and we predict $T=t$ or $T>t$ based on $g(\bZ(t))$, with a larger value being more indicative of $T=t$. Following \cite{heagerty2005survival}, the false positive rate is $\FPR_t(c) = P\{g(\bZ(t))>c\mid T > t\} $, the true positive rate is $\TPR_t(c) = P\{g(\bZ(t))>c\mid T=t\}$, and the ROC function is $\ROC_t(q) = \TPR_t(\FPR_t^{-1}(q))$. It is known that, when predicting a binary disease outcome with multiple disease markers, the risk score (i.e., the probability of disease given markers) yields the highest ROC curve \citep{mcintosh2002combining}. For survival outcomes, the hazard $\lambda(t\mid \bZ(t))$ can be viewed as an analog of the risk score. Following arguments of the Neyman-Pearson Lemma, setting $g(\cdot) = \lambda(t\mid \cdot)$ yields the highest $\ROC_t$. Thus $\widehat\lambda_{\T}(t\mid\cdot)$ with a higher $\ROC_t$ curve is desired. }However, when evaluating discrete-valued markers such as a tree-based risk score, the $\ROC_t(\cdot)$ degenerates to a finite number of points. Hence important summary measures such as AUC are not well defined. We fill in the gap by introducing a generalized time-dependent ROC curve.

\subsection{Generalized ROC curves for evaluating survival trees}
With a finite number of terminal nodes, $\widehat\lambda_{\T}(t\mid \bz)$ at a fixed time $t$ is a discrete-valued scalar function of $\bz$. Thus the $\ROC_t$ function for $\widehat\lambda_{\T}(t\mid \bZ(t))$ becomes a finite set of points rather than a continuous curve. More generally, if $g(\bZ(t))$ has a point mass at $c$, the function  $\ROC_t(q)$  is undefined for $q\in(\FPR_t(c),\FPR_t(c-))$, where $\FPR_t(c-) = \lim_{a\rightarrow c^-}\FPR_t(a)$. In this case, we construct a continuous curve, denoted by $\ROC^*_t$, via linear interpolation. Specifically, for $q\in(\FPR_t(c),\FPR_t(c-))$, the point $(q,\ROC_t^*(q))$ on the $\ROC^*_t$ curve corresponds to the following prediction rule: if $g(\bZ(t))>c$, predict $T=t$; if $g(\bZ(t)) = c$, predict $T=t$ with probability $\{q-\FPR_t(c)\}/\{ \FPR_t(c-)-\FPR_t(c)\}$; and if $g(\bZ(t)) < c$, predict $T>t$. In the special case where $g(\bZ(t))$ is a continuous variable, $\ROC^*_t$ reduces to $\ROC_t$. The optimality of true hazard function $\lambda(t\mid \cdot)$ with respect to $\ROC^*_t$ is established in Proposition \ref{p22}.  Our result suggests that $\ROC^*_t$ can be used to evaluate the predictive ability of $\widehat\lambda_{\T}(t\mid\cdot)$, and a higher $\ROC^*_t$ curve is favorable.  The mathematical definition of $\ROC^*_t$ and the proof of Proposition 1 is given in the Supporting Information. 

\begin{proposition}[Optimality of the hazard function]
	\label{p22}
	{At time $t$, among all scalar functions $g:\mathcal{Z}_t\mapsto\mathbb{R}$, the hazard $\lambda(t\mid \cdot)$ defined in (\ref{haz}) is optimal in the sense that $g(\bZ(t)) = \lambda(t\mid \bZ(t))$ yields the highest $\ROC_t^*$ curve.}
\end{proposition}

The area under the $\ROC^*_t$ curve, defined as $\AUC^*_t = \int_0^1 \ROC^*_t(q)\dev q$, has the interpretation of a concordance measure \citep{pepe2003statistical}. {It can be shown that $\AUC^*_t$ is equivalent to}
\begin{align*}
\CON_t(g)
=~ P\{g(\bZ_1(t))>g(\bZ_2(t)) \mid T_2>T_1 = t\}+ \frac{1}{2}P\{g(\bZ_1(t))=g(\bZ_2(t)) \mid T_2>T_1 = t\},
\end{align*}
where $\{\bZ_i(\cdot),T_i \}, i = 1, 2,$ are i.i.d. replicates of $\{\bZ(\cdot),T \}$. Based on Proposition \ref{p22}, $\CON_t$ can be used to evaluate the predictive ability of $g(\bZ(t))$. We then consider a global measure to evaluate $\widehat{\lambda}_{\T}$ on $[0,s]$. Define a function $\widetilde{g}:\mathbb{R}^+\times \mathcal{Z}_t\mapsto \mathbb{R}$ that combines $\bZ(t)$ in a time-dependent way and use $\widetilde{g}(t,\bZ(t))$ to characterize the risk at $t$ among survivors at $t$. We then integrate $\CON_t(\widetilde{g}(t,\cdot))$ over $t$ with a weight function $\omega(t)$ and define,
$$\ICON(\widetilde{g}) = \int_{0}^{s} \omega(t)\CON_t(\widetilde{g}(t,\cdot))\dev t.$$
In practice, the weight functions $\omega(t)$ can be specified by study investigators. A simple example is to set $\omega(\cdot) = 1$. Another possible choice is to set $\omega(t) = f(t)S(t)/P(T_2 > T_1, T_1<s )$, where $f(t),S(t)$ are the marginal density and survival functions of $T$, respectively. Then we have $\ICON(\widetilde{g}) = P\{\widetilde{g}(T_1,\bZ_1(T_1))>\widetilde{g}(T_1,\bZ_2(T_1)) \mid T_2 > T_1, T_1< s\} + 0.5 P\{\widetilde{g}(T_1,\bZ_1(T_1))=\widetilde{g}(T_1,\bZ_2(T_1)) \mid T_2 > T_1, T_1< s \},$
which is the probability that the subject failing earlier has a higher risk at the failure time. Following Proposition \ref{p22}, the true hazard $\lambda$ maximizes $\ICON$. Motivated by this fact, we propose to use $\ICON$ as a guidance to build survival trees.

\begin{remark}
	The Harrell's C-statistic \citep{harrell1982evaluating} has been commonly used to quantify the capacity of a risk score at baseline in discriminating among subjects with different event risks. When the event time is subject to censoring, the population parameter corresponding to the Harrell's C depends on the censoring distribution. \cite{uno2011c} studied a C-statistic that is consistent for a censoring free population concordance measure under a working Cox model $\lambda(t\mid \bX) = \lambda_0(t)\exp(\bX\bbeta)$, and $\bX\bbeta$ maximizes the limiting value of the Uno's C-statistic. However, without the Cox model assumption, it is not clear how to combine the predictors so that the limiting value is maximized. The proposed ICON is maximized when $\widetilde{g}(t,\bz) = \lambda(t\mid \bz)$ and is proper for guiding the tree building procedure.
\end{remark}

\vspace{-0.2in}

\subsection{Splitting and pruning based on ICON}
We next develop an algorithm to construct a time-invariant partition. Although the assumption (A1) is adopted for establishing the large-sample properties of the tree-based estimation, stronger assumptions are often generally to understand the splitting criteria, especially in the early steps of splitting. For example, when selecting the optimal split at the root node, the log-rank splitting rule implicitly assumes that $C$ is independent of $T$ within the child nodes, which is not guaranteed by assumption (A1). For ease of discussion, we assume  $C$ is independent of $\{ T, \bZ(\cdot)\}$ in Section 3.2. An extension to handle covariate-dependent censoring is given in Section 3 of the Supporting Information.

We first consider the estimation of $\ICON(\lambda_{\T})$ using the training data. For a node $\tau$, define $S(t,\tau) = P\{ \bZ(t)\in\tau, T\ge t \}$ and $f(t,\tau) = \lim_{\delta_t\rightarrow 0^+}P\{T\in[t,t+\delta_t) , \bZ(t)\in\tau \}/\delta_t$. Given a partition  $\tT= \{\tau_1,\tau_2,\ldots,\tau_M \}$, the estimation of $\CON_t(\lambda_{\T}(t\mid \cdot)) $ is developed from
\begin{equation*}
\CON_t(\lambda_{\T}(t\mid \cdot)) =  \frac{\sum_{j=1}^M\sum_{k=1}^M I\{\lambda(t\mid\tau_j)>\lambda(t\mid\tau_k)\}f^{}(t,\tau_j)S^{}(t,\tau_k)+ 0.5\sum_{j=1}^M f^{}(t,\tau_j)S^{}(t,\tau_j)}{\sum_{j=1}^M \sum_{k=1}^M  f^{}(t,\tau_j)S^{}(t,\tau_k)}.
\end{equation*}
Under independent censoring, we have $f(t,\tau) = f^{\ast}(t,\tau)/P(C\ge t)$ and $S(t,\tau) = S^{\ast}(t,\tau)/P(C\ge t)$.
Therefore, a consistent estimator for the concordance measure is given by 
\begin{align}
\label{contest}\nonumber
\widehat{{\CON}}_t&(\widehat{\lambda}_{\T}(t\mid \cdot)) =\\
&\frac{\sum_{j=1}^M\sum_{k=1}^M I\{\widehat{\lambda}(t\mid\tau_j)>\widehat{\lambda}(t\mid\tau_k)\}\widehat{f}^{\ast}(t,\tau_j)\widehat{S}^{\ast}(t,\tau_k)+ 0.5\sum_{j=1}^M \widehat{f}^{\ast}(t,\tau_j)\widehat{S}^{\ast}(t,\tau_j)}{\sum_{j=1}^M \sum_{k=1}^M  \widehat{f}^{\ast}(t,\tau_j)\widehat{S}^{\ast}(t,\tau_k)}.
\end{align}
Note that the usual $O(n^2)$ computational costs for a concordance measure can be reduced by using (\ref{contest}), because the tree-based hazard ${\lambda}_{\T}(t \mid \bz)$ at time $t$ takes $M$ discrete values. To estimate ${\ICON}$, we use 
$\widehat{\ICON}(\widehat{\lambda}_{\T})= \int_{0}^{{s}}  \widehat{\CON}_t(\widehat{\lambda}_{\T}(t\mid \cdot))\widehat{\omega}(t)\dev t$,
where $\widehat{\omega}(t)$ is a weight function that possibly depends on the data. In practice, one can approximate the integral by the trapezoidal rule. Alternatively, one can also define ICON as a weighted average of $\CON_t$ on discrete time points. As demonstrated in our simulation studies, a moderate number of time points (e.g., 20) can yield reasonably good performances.

To build a partition, we begin at the top of the tree and then successively split the predictor space. {At each splitting step, an optimal split can be chosen according to certain criteria.} Consider the partition $\tT = \{\tau_1,\ldots,\tau_M \}$ and a split on any arbitrary node in $\tT$. Without loss of generality, suppose $\tau_1$ is split into $\tau_1^L$ and $\tau_1^R$, and the partition after splitting is denoted by $\T' = \{\tau_1^L,\tau_1^R, \tau_m; m = 2,\ldots,M \}$. Proposition \ref{split2} shows that splitting increases the concordance as long as the hazards of the two child nodes are different. The proof is given in the Supporting Information.

\begin{proposition}[Splitting increases $\CON_t$ when two child nodes are heterogeneous]
	\label{split2} 
	Let $\lambda_{\T}$ and $\lambda_{\T'}$ be partition-based hazard functions before and after splitting, respectively. Then 
	$\CON_t(\lambda_{\T'}(t\mid\cdot)) \ge \CON_t(\lambda_{\T}(t\mid\cdot))$, 
	and the equality holds if and only if $\lambda(t\mid\tau_1) = \lambda(t\mid\tau_1^L) = \lambda(t\mid\tau_1^R)$. Moreover, we use ``$\widehat{\hspace{1em}}$'' to denote the estimated values using (\ref{contest}), then
	$\widehat{\CON}_t(\widehat{\lambda}_{\T'}(t\mid\cdot)) \ge \widehat{\CON}_t(\widehat{\lambda}_{\T}(t\mid\cdot))$, and the equality holds if and only if $\widehat{\lambda}(t\mid\tau_1) = \widehat{\lambda}(t\mid\tau_1^L) = \widehat{\lambda}(t\mid\tau_1^R)$.
\end{proposition}
Based on Proposition \ref{split2}, we have $\widehat{\ICON}(\widehat{\lambda}_{\T'}) \ge \widehat{\ICON}(\widehat{\lambda}_{\T})$. When {$\widehat{\omega}(\cdot)>0$}, the equality holds if and only if $\widehat{\lambda}(\cdot\mid\tau_1^L) = \widehat{\lambda}(\cdot\mid\tau_1^R)$ almost everywhere on $(0,s]$. 
The validity of Proposition \ref{split2} does not depend on the censoring distribution. In practice,  $\widehat{\ICON}(\widehat \lambda_{\T})$ may not correctly estimate ${\ICON}(\lambda_{\T})$ if the independent censoring assumption is violated, but the  true concordance usually increases after splitting. 

As ICON can effectively detect the difference in hazards of the two child nodes, it is natural to consider choosing the optimal split that maximizes ICON. However, an ICON-based splitting rule is non-local because the split depends on the data in the parent node as well as other parts of the tree. As a result, additional computational burdens can arise when the number of terminal nodes in the current tree is large. In what follows, we introduce a local splitting rule, where the optimal split is chosen to maximize the increment of ICON within a node. For node $\tau$'s child nodes $\tau^L$ and $\tau^R$, it can be shown that  $\CON_t$ within $\tau$ is 0.5 before splitting and is 
$0.5+ 0.5{|f(t, \tau^L)S(t, \tau^R)-f(t, \tau^R)S(t, \tau^L)|}/{f(t, \tau)S(t, \tau)}$
after splitting. Hence the increment in ICON within $\tau$ after splitting is
\begin{equation*}
\Delta\ICON_{\tau} =\int_{0}^{s} \frac{|f(t, \tau^L)S(t, \tau^R)-f(t, \tau^R)S(t, \tau^L)|}{f(t, \tau)S(t, \tau)}\omega(t)\dev t,
\end{equation*}
which can be estimated by 
\begin{align*}
\widehat{\Delta\ICON_{\tau}} =\int_{0}^{s} \frac{|\widehat{f}^{\ast}(t,\tau^L)\widehat{S}^{\ast}(t,\tau^R) -\widehat{f}^{\ast}(t,\tau^R)\widehat{S}^{\ast}(t,\tau^L)|}{\widehat{f}^{\ast}(t,\tau)\widehat{S}^{\ast}(t,\tau)}\widehat{\omega}(t)\dev t.
\end{align*}

Although splitting generally increases the concordance, a large tree can overfit the data, and the survival estimate within a small terminal node could be biased. Similar to the CART algorithm, we continue splitting until a pre-determined stopping criterion on the minimal node size is met and then prune the fully grown tree.  A node $\tau$ is considered to be ``splittable'' only if $n(\tau) \overset{\rm def}{=} \sum_{i=1}^nI(\bZ_i(0)\in\tau)\ge n_{\min}$, where $n_{\min}$ is a pre-specified constant. We also require all the nodes in the tree to satisfy that $n(\tau)\ge n_{\min}/2$. After the stopping criterion is met, we use the following concordance-complexity measure for pruning,
$\ICON_{\alpha}(\T) = \widehat{\ICON}(\widehat \lambda_{\T})-\alpha|\T|,$
where $|{\tT}| $ is the number of terminal nodes in $\tT$ and $\alpha$ is a complexity parameter. For a fixed $\alpha$, a tree with a large value of $\ICON_{\alpha}(\T)$ is generally favorable. Let $K$ be the number of terminal nodes in the un-pruned tree. 
For $k = 1, \ldots, K$, let $\T_{(k)}$ be the size-$k$ subtree that has the largest value of $\widehat{\ICON}$. For each $\alpha$, we define the optimal subtree $\T^\alpha$ as the subtree that has the largest $\ICON_{\alpha}$ among $\mathcal{C} \overset{\rm def}{=} \{\T_{(1)},\ldots,\T_{(K)}  \}$. For $\alpha_0 = 0$, the tree $\T^{\alpha_0} = \T_{(K)}$ is the optimal subtree. Define $\alpha_{\T,\T'} = \{\widehat{\ICON}(\widehat{\lambda}_{\T'})-\widehat{\ICON}(\widehat{\lambda}_{\T})\}/(|\tT'|-|\tT|)$. The $q$th $(q\ge 1)$ threshold parameter $\alpha_q$ is defined as $\alpha_q= \min\{\alpha_{\T,\T^{\alpha_{q-1}}}; |\tT| \le |\tT^{\alpha_{q-1}}|,\T\in \mathcal{C}\}$ and $\T^{\alpha_q}$ is defined as the smallest tree in $\{\T\mid \alpha_{\T,\T^{\alpha_{q-1}}} =\alpha_q, |\tT| \le |\tT^{\alpha_{q-1}}|,\T\in \mathcal{C} \}$. Note that there exists an integer $Q$ such that $\T^{\alpha_Q} = \T_{(1)}$. For $\alpha\in[\alpha_q,\alpha_{q+1})$ and $q<Q$, $\T^{\alpha_q}$ is the optimal subtree; and for $\alpha\in[\alpha_Q,\infty)$, $\T^{\alpha_Q}$ is the optimal subtree. In practice, $\alpha$ and $\T^{\alpha_Q}$ can be determined by cross-validation. The survival tree algorithm is given in Algorithm 1 outlined in Table \ref{tab:a1}.

\begin{table}
	\caption{The survival tree algorithm}
	\label{tab:a1}
\begin{algorithm}[H]
\caption{The ROC-guided survival tree algorithm}
\SetKwInOut{Input}{Input}
\SetKwInOut{Output}{Output}

\Input{The training data are $\{Y_i,\Delta_i,\bZ_i^H(Y_i); i = 1,\ldots,n \}$. }
\Output{ 
A time-invariant partition that can be used for risk prediction.}
\textbf{Splitting} \\
Start from the root node, which is labeled as node 1. Let $m$ denote the number of nodes (including both internal and terminal nodes) in the current tree and $m = 1$. Set $k=1$. 

\While{$k \le m$}
{
 \If{node $k$ is splittable}{
 Identify all possible splits on node $k$\;
 Find the split that results in the largest $\widehat{\Delta\ICON_\tau}$\;
 Update the tree with the selected split and label the left and right children nodes to be node $m+1$ and $m+2$, respectively\;
 Set non-splittable children nodes as terminal nodes\;
 $m \leftarrow m + 2$\;}
 $k \leftarrow k + 1$\;
}

\textbf{Pruning}\\
Calculate $\alpha_q$ and identify $\T^{\alpha_q}$ for $q = 1,\ldots,Q$. For $q<Q$, set $\beta_q = \sqrt{\alpha_q\alpha_{q+1}}$ as the representative value of the interval $[\alpha_q,\alpha_{q+1})$ and $\beta_Q=\alpha_Q$\;

Select the optimal $\beta_q$ and the corresponding tree using cross-validation.
\end{algorithm}
\end{table}

\begin{remark}
	In practice, one may consider node-specific bandwidths such that the bandwidth for node $\tau$ is $h_{\tau} = cn_{\tau}^{-1/5}$, where $n_\tau = \sum_{i=1}^n I(\bZ_i(Y_i)\in\tau)$. An order of $n_{\tau}^{-1/5}$ is chosen to achieve the lowest order of integrated mean square error within $\tau$. Specifically, following the arguments of existing works on smoothing hazard, it can be shown that, $\int_{0}^{s}E\{\widehat{\lambda}(t\mid \tau) -   {\lambda}(t\mid \tau)\}^2 dt  = O(h_{\tau}^4 + n_\tau^{-1} h_\tau^{-1}) = O(n_\tau^{-4/5})$. An ad hoc choice of $c$ is $c_0 = s/8$ \citep{muller1994hazard}. In practice, one can also choose $c$ via cross-validation. 
\end{remark}

\vspace{-.4in}

\section{Survival ensembles based on martingale estimating equations}

Survival trees can be transformed into powerful risk prediction tools by applying ensemble methods such as bagging \citep{breiman1996bagging} and random forests \citep{breiman2001random}. The idea in bagging is to average many noisy but approximately unbiased tree models to reduce the variance; random forests further improve the variance by reducing the correlation between the trees via a random selection of predictors in the tree-growing process. In random forests for regression and classification \citep{breiman2001random}, the prediction for a new data point is the averaged prediction from individual trees that are often grown sufficiently deep to achieve low bias. For right-censored data, when the sizes of terminal nodes are very small, the within-node estimates of survival or cumulative hazard functions could be biased \citep{chen1982small,pena1993small}. In what follows, we propose to average the unbiased martingale estimating equations rather than directly averaging node summaries. We treat forests as a type of adaptive nearest neighbor estimator \citep{lin2006random, meinshausen2006quantile, athey2018generalized} and propose local estimation for the survival or hazard functions.

Let $\mathbb{T} = \{\T_b\}_{b = 1}^B$ be a collection of $B$ partitions constructed using bootstrap samples. Each partition is constructed via a splitting procedure where at each split, $m~(m<p)$ predictors are randomly selected as candidates for splitting. The splitting criterion can either be the concordance measures in Section 3 or other appropriate criteria. Given $\T_b$ and a terminal node $\tau\in\T_b$, one can solve the following estimating equation for the node-specific hazard at $t$,
$\sum_{i=1}^n w_{bi}{I\left(\bZ_i(t) \in \tau \right)} \{\dev N_i(t)-I(Y_i\ge t)\lambda(t\mid\tau) \dev t \}=0$, where $w_{bi}\in \mathbb{N}_0$ is the frequency of the $i$th training observation in the $b$th bootstrap sample. Let $l_{\T_b}\{\bz\}$ be the partition function for $\T_b$ so that $l_{\T_b}\{\bz\} = \tau$ if and only if $\bz\in\tau$ and $\tau\in\mT_b$, then the $b$th tree induces the following estimating equation for $\lambda(t\mid \bz)$, 
\begin{align}
\label{eeb}
\sum_{i=1}^n w_{bi}{I\left(\bZ_i(t) \in l_{\T_b}\{\bz\} \right)} \{\dev N_i(t)-I(Y_i\ge t)\lambda(t\mid \bz)\dev t \}=0.
\end{align}
To get ensemble-based prediction, we take average of the estimating functions in (\ref{eeb}) from all the $B$ trees and obtain the following local martingale estimating equation for $\lambda(t\mid \bz)$, 
\begin{align}
\label{ee}
\sum_{i=1}^n w_{i}(t,\bz)\{\dev N_i(t)-I(Y_i\ge t)\lambda(t\mid \bz) \dev t \}=0,
\end{align}
where the weight function is $w_i(t, \bz) =\sum_{b=1}^B w_{bi}{I\left(\bZ_i(t) \in l_{\T_b}\{\bz\} \right)}/B$.
The weight $w_i(t,\bz)$ captures the frequency with which the $i$th observation $\bZ_i(t)$ falls into the same node as $\bz$. Based on Equation (\ref{ee}), a kernel type estimator for the hazard function $\lambda(t\mid \bz)$ is given by
\begin{align}
\label{4rstHaz}
\widehat{\lambda}_{\mathbb{T}}(t\mid\bz) =\int_{0}^{\infty} K_h(t-u) \frac{ \sum_{i=1}^{n}w_i(u,\bz)\dev N_i(u)}{ \sum_{i=1}^{n}w_i(u,\bz)I(Y_i\ge u)} .
\end{align}
Boundary correction is the same as that of survival trees. When $\bZ(t)$ are external time-dependent covariates and the hazard at $t$ depend on $\bZ^H(t)$ only through $\bZ(t)$, the survival probability at $t$ given $\bZ_0^H(t) = \{\bZ_0(u), 0\le u\le t \}$  is predicted to be
\begin{align}
\exp\left\{ -\int_{0}^t \frac{\sum_{i=1}^{n}w_i(u,\bZ_0(u))\dev N_i(u)}{\sum_{i=1}^{n}w_i(u,\bZ_0(u))I(Y_i\ge u)} \right\}.
  \label{4rstSurv}
\end{align}
Details of the prediction procedure are given in Algorithm 2 outlined in Table \ref{tab:a2}.

The algorithm can also be extended to incorporate subsampling without replacement and sample-splitting  \citep{athey2018generalized}.  Specifically, one can divide the $b$th subsample from the original training data into two halves, whose indices are denoted by $\mathcal{I}_{1b}$ and $\mathcal{I}_{2b}$. Using the ${\mathcal I}_{1b}$ sample to place the splits and holding out the ${\mathcal I}_{2b}$ sample for within-leaf estimation yields honest trees. The honesty condition is proven to be successful in the literature on regression forests; readers are referred to \citep{wager2017estimation} for an in-depth discussion. With subsampling and sample-splitting, the weight can be calculated as before but with $w_{bi}$ being the frequency of the $i$th observation in $\mathcal{I}_{2b}$.

\begin{table}
	\caption{The survival ensemble algorithm}
	\label{tab:a2}
\begin{algorithm}[H]
\SetAlgoLined
\SetKwInOut{Input}{Input}
\SetKwInOut{Output}{Output}

		\caption{The survival ensemble algorithm}
		\Input{The training data are $\{Y_i,\Delta_i,\bZ_i^H(Y_i); i = 1,\ldots,n \}$. }
		\Output{
		 The predicted hazard at $t$ given $\bz_t$ or the predicted survival probability at $t$ given covariate history $\{\bz_u, 0\le u\le t\}$.}
		
		In what follows, define $|\mathcal{N}_{b}(u,\bz)| = \sum_{i=1}^n w_{bi}I(Y_i\ge u, \bZ_i(u)\in l_{\T_b}\{\bz\} )$, where $w_{bi}\in \mathbb{N}_0$ is the frequency of the $i$th training observation in the $b$th bootstrap sample. 

		\bigskip
		
		\textbf{Hazard prediction}
		
		\noindent {Initialize the weights: $ (v_{01},\ldots,v_{0n}) =  (v_{11},\ldots,v_{1n}) = (0,0,\ldots,0)$.}		
		
				~ \For{$b = 1$ to $B$}{
			Draw the $b$th bootstrap sample from the training data\;
			
			Construct a partition $\T_b$ using the $b$th bootstrap sample with a random selection of $m$ features at each split without pruning\;

			~ \For{$i = 1$ to $n$}{
				\If{$\bZ_i(Y_i)\in l_{\T_b}\{\bz_t\}$}{$v_{0i} \leftarrow v_{0i} + \Delta_i w_{bi}$\;
				}
				$v_{1i} \leftarrow v_{1i} + |\mathcal{N}_{b}(Y_i,\bz_t)|$\;
				}
		
		}

		Predict the hazard at $t$ given $\bz_t$ using 
		$\sum_{i=1}^{n} K_h(t-Y_i)v_{0i} / v_{1i}$.
		\bigskip
		
		\textbf{Survival probability prediction}
		
		\noindent {Initialize the weights: $ (v_{21},v_{22},\ldots,v_{2n}) =  (v_{31},v_{32},\ldots,v_{3n}) = (0,0,\ldots,0)$.}

		~ \For{$b = 1$ to $B$}{
			Draw the $b$th bootstrap sample from the training data\;

			Construct a partition $\T_b$ using the $b$th bootstrap sample with a random selection of $m$ features at each split without pruning\;

		~ \For{$i = 1$ to $n$}
		{
				\If{$\bZ_i(Y_i)\in l_{\T_b}\{\bz_{Y_i}\}$}
				{
				$v_{2i} \leftarrow v_{2i} + \Delta_i w_{bi}$\;
				}
		$v_{3i} \leftarrow v_{3i} +  |\mathcal{N}_{b}(Y_i,\bz_{Y_i})|$\;
		}
		
	}		
			
		 Predict the survival probability at $t$ given $\{\bz_u, 0\le u\le t\}$ using 
		$\exp\{-\sum_{i=1}^{n} I(Y_i \le t) v_{2i} / v_{3i}\}$.

\end{algorithm}
\end{table}

\vspace{-1cm}
\section{Simulation Studies}
\label{sect:simu}
We conducted extensive simulation studies to investigate the performance of the proposed methods. Our simulation settings are described below:
\begin{description}
\item[(\RomanI)] The survival times follow an exponential distribution with mean $\exp\left(0.1 \sum_{j = 11}^{25}Z_j\right)$, where the covariate vector $(Z_1, \ldots, Z_{25})$ follows a multivariate normal distribution with mean zero and a covariance matrix with the $(i, j)$th element equal to $0.9^{|i - j|}$. The censoring times follow an independent exponential distribution with mean $\eta_1$.
\item[(\RomanII)] The survival times follow an exponential distribution with mean 
$\sin(Z_1\pi) + 2|Z_2 - 0.5| + Z_3^3$, where  the elements in the covariate vector $(Z_1, \ldots, Z_{25})$ are i.i.d. uniform random 
variables on $[0, 1]$. The censoring times follow an independent uniform distribution on $[0, \eta_2]$.
\item[(\RomanIII)] The survival times followed a gamma distribution with shape parameter $0.5 + 0.3 |\sum_{j = 11}^{15}Z_j|$
and scale parameter 2, where the covariate vector $(Z_1, \ldots, Z_{25})$ follows a multivariate normal with mean zero and a covariance
matrix with the $(i, j)$th element equal to $0.75^{|i - j|}$. 
The censoring times follow an independent uniform distribution on $[0, \eta_3]$.
\item[(\RomanIV)] The survival times follow a log-normal distribution with mean $\mu=0.1|\sum_{j = 1}^5Z_j| + 0.1|\sum_{j = 21}^{25}Z_j|$ and scale parameter one, 
where the covariate vector $(Z_1, \ldots, Z_{25})$ is multivariate normal with mean zero and a covariance matrix with the $(i, j)$th element equal to $0.75^{|i - j|}$. 
The censoring times follow a log-normal distribution with mean $\mu + \eta_4$ and scale parameter one. 
\item[(\RomanV)] The hazard function is 
$\lambda(t\mid \bZ(t)) = \exp\{0.5\sum_{j = 1}^{10}Z_j(t) + Z_{11}\}/10$, 
where the time-dependent covariates $\{Z_1(t), \ldots, Z_{10}(t)\}$ are multivariate 
normal with mean $kt + b$ and a covariance matrix with the $(i, j)$th element equal to $0.9^{|i - j|}$ at $t$, 
and the covariate $\{Z_{11}, \ldots, Z_{20}\}$ is uniform on [0, 1].
The censoring times follow a uniform distribution on $[0, \eta_5]$.
\item[(\RomanVI)] The hazard function is
$\lambda(t \mid \bZ(t)) = \sum_{j = 1}^{10}\{Z_{j}(t)-Z_{j + 10}\}^2$. The time-dependent covariates are set as $Z_j(t) = tk_j/10$, with $k_j$ drawn uniformly from $[0,1]$. The covariates $(Z_{11}, \ldots, Z_{20})$ are i.i.d. uniform random variables on $[0, 1]$.
The censoring times follow a uniform distribution on $[0, \eta_6]$.
\item[(\RomanVII)] The hazard function is
$\lambda(t \mid \bZ(t)) = \sum_{j = 1}^{10}\{Z_{j}(t)-Z_{j + 10}\}^2$. The time-dependent covariates $Z_j(t) = tk_j/10$ and $(k_{1}, \ldots, k_{10})$ are generated from a multivariate normal distribution
with mean one and a covariance matrix with the $(i, j)$th element equal to $0.9^{|i - j|}$. The covariates $(Z_{11}, \ldots, Z_{20})$ are i.i.d. standard normal random variables.
The censoring times follow a uniform distribution on $[0, \eta_7]$.
\end{description}
Scenarios with only time-independent covariates were motivated from the settings in \cite{zhu2012recursively} and \cite{steingrimsson2018censoring}. The other three scenarios were included to examine the proposed methods in dealing with time-dependent covariates. For each setting, the censoring parameter $\eta_j$ is tuned to yield censoring percentages of $0\%$, $25\%$, and $50\%$. The censoring distribution depends on covariates in Scenario (\RomanIV). The proportional hazards assumption holds in Scenarios (\RomanI) and (\RomanV) but is violated in other scenarios. 

We explored splitting based on the increment in within node ICON and overall ICON (i.e., $\Delta\widehat{\ICON}_\tau$ in Algorithm 1 was replaced with the estimated ICON of the whole tree). All the covariates were transformed through $\widehat{F}_t(\cdot)$ as in Remark 1 and treated as time-dependent covariates. For the ROC-guided survival trees, ten-fold cross-validation was used to choose the tuning parameter $\alpha$ in the concordance-complexity measure. For the ensemble method, $500$ unpruned survival trees were constructed using bootstrap samples, and a set of randomly selected $\left \lceil \sqrt p \right \rceil$ features were considered at each split. For both methods, the minimum number of baseline observations in each terminal node was set at 15. The Epanechnikov kernel function $K(x) = 0.75(1 - x^2)I(|x| \le1)$ was used, where the bandwidth $h=t_0/20$ for $t_0$ equals the 0.95 quantile of the uncensored survival times. The ICON measure was chosen as the average of $\CON_t$ on 20 equally spaced quantiles of the uncensored survival times.

The proposed methods were compared to several available implementations including the random survival forest (RSF) \citep{ishwaran2010consistency}, the relative risk tree \citep{leblanc1992relative}, and the Cox proportional hazards model, implemented in the \texttt{R} packages \texttt{randomForestSRC} \citep{rfsrc}, \texttt{rpart} \citep{rpart}, and \texttt{survival} \citep{survival-package}, respectively. The RSFs were fitted using the default settings except that 500 trees were used in the ensembles. In addition to the default node size of 15, we also included an RSF algorithm using the node size that gives the smallest out-of-bag error. Ten-fold cross-validation was used for pruning in relative risk trees. To the best of our knowledge, there are no existing software on survival trees and forests that can predict the survival probability based on covariate history, so both the RSF and the relative risk tree algorithms were fitted using the baseline values of all covariates.

The integrated absolute error (IAE) is used to evaluate the prediction accuracy of different methods. For each fitted model, we predicted the survival probabilities for $n_0=500$ new observations $\{ \bZ^{new}_i(u), u \ge 0 , i=1,\ldots,n_0\}$  generated from the distribution of the training data. When there are only time-independent covariates, the IAE is defined as 
$(n_0s)^{-1}\sum_{i=1}^{n_0}\int_0^{s} \left|\widehat{P}\{T^{new}_i\ge t| \bZ_i^{new}(0) \} - P\{T^{new}_i\ge t|\bZ_i^{new}(0)\}\right|\, \dev t$,
where $s$ is the 0.95 quantile of the survival time. In the presence of time-dependent covariates, the IAE is defined as 
$(n_0s)^{-1}\sum_{i=1}^{n_0}\int_{0}^{s} \left| \hP\{T_i^{new}\ge t\mid  \bZ^{new,H}_i(t)\}-P\{T_i^{new}\ge t\mid  \bZ^{new,H}_i(t)\} \right|\dev t$, where $\bZ^{new,H}_i(t) = \{ \bZ^{new}_i(u), 0\le u \le t \}$. For the proposed methods, the predicted survival probabilities were obtained from~\eqref{spred} and~\eqref{4rstSurv}, with covariates transformed through $\widehat{F}_t(\cdot)$.

Table~\ref{tab:sim} reports the summary statistics of the simulation results for sample sizes $n = 200$ and $1000$. The results for $n = 100$ and 500, and the average terminal node sizes in ROC-guided trees after pruning are presented in the Supporting Information. The two splitting rules in the proposed methods yield similar IAEs for both trees and ensembles. Thus we recommend the use of the $\Delta\widehat{\ICON}_\tau$ rule due to its lower computational cost. As expected, the proposed ensemble method has the lowest IAE in most scenarios, and the error decreases as $n$ increases. Interestingly, with a small to moderate sample size (e.g., $n\le200$), our ensemble method outperforms the Cox model when simulating under the proportional hazard assumption of Scenarios (\RomanI) and (\RomanV). In the absence of time-dependent covariates, the proposed ensemble method and RSF generally perform better than single survival trees. On the other hand, the ROC-guided tree and the relative risk tree have similar performances. In the presence of time-dependent covariates, the proposed ensemble method continues to outperform all the other methods except for the Cox model when $n = 1000$ in Scenario (\RomanV). The ROC-guided tree also performs well but has larger IAEs than the Cox model in Scenario (\RomanV) when the proportional hazard assumption holds. The simulation results indicate that the proposed methods perform well and incorporating time-dependent covariates improves the prediction accuracy substantially.

\begin{sidewaystable}[ht]
  \caption{Summaries of integrated absolute errors ($\times1,000$).
    The numbers 0\%, 25\%, 50\% correspond to the different censoring proportions.
    ROC-Tree and ROC-Ensemble are the proposed ROC-guided survival trees and ensembles
    with either the $\Delta$ICON or the ICON as the splitting criterion.
    RSF is the random survival forests methods implemented  in \texttt{R} package \texttt{randomForestSRC}
    with either the default settings or with the \texttt{nodesize} parameter optimized
    based on out-of-bag error (using the function \texttt{tune.nodesize}).
    The Cox regression is implemented in \texttt{R} package \texttt{survival}.
    RR-Tree is the relative risk tree implemented in \texttt{R} package \texttt{rpart}.
  }
  \label{tab:sim}
  \centering
  \renewcommand\tabcolsep{3.2pt}
  \renewcommand{\arraystretch}{1}
  \begin{tabular}{rr rrrrrr rrrrrr rrrrrr rrr rrr}
    \toprule
    && \multicolumn{6}{c}{ROC-Tree} & \multicolumn{6}{c}{ROC-Ensemble}  & \multicolumn{6}{c}{RSF} \\
    \cmidrule(r){2-7}\cmidrule(r){8-13}\cmidrule(r){14-19}
    & \multicolumn{3}{c}{$\Delta$ICON} & \multicolumn{3}{c}{ICON} & \multicolumn{3}{c}{$\Delta$ICON} & \multicolumn{3}{c}{ICON}
                                       & \multicolumn{3}{c}{default} & \multicolumn{3}{c}{optimal} & \multicolumn{3}{c}{Cox} & \multicolumn{3}{c}{RR-Tree} \\
    \cmidrule(r){2-4}\cmidrule(r){5-7}\cmidrule(r){8-10}\cmidrule(r){11-13}\cmidrule(r){14-16}\cmidrule(r){17-19}\cmidrule(r){20-22}\cmidrule(r){23-25}
    Sce & 0\% & 25\% & 50\%  & 0\% & 25\% & 50\%  & 0\% & 25\% & 50\% & 0\% & 25\% & 50\% & 0\% & 25\% & 50\% & 0\% & 25\% & 50\% & 0\% & 25\% & 50\% & 0\% & 25\% & 50\% \\
    \midrule
    & \multicolumn{24}{c}{Scenarios with time-independent covariates}\\
    & \multicolumn{24}{c}{$n = 200$}\\
    \RomanI & 82 & 95 & 152 & 81 & 95 & 151 & 49 & 54 & 89 & 49 & 54 & 89 & 59 & 70 & 126 & 74 & 87 & 134 & 48 & 57 & 99 & 87 & 101 & 146 \\ 
    \RomanII & 103 & 107 & 166 & 104 & 108 & 165 & 77 & 79 & 124 & 77 & 79 & 124 & 75 & 79 & 130 & 90 & 101 & 161 & 110 & 118 & 175 & 82 & 84 & 131 \\ 
    \RomanIII & 121 & 127 & 175 & 120 & 127 & 176 & 101 & 103 & 135 & 100 & 103 & 134 & 99 & 102 & 136 & 119 & 131 & 175 & 152 & 158 & 196 & 122 & 125 & 160 \\ 
    \RomanIV & 92 & 100 & 127 & 92 & 100 & 126 & 53 & 58 & 70 & 53 & 58 & 70 & 57 & 67 & 96 & 71 & 83 & 111 & 76 & 82 & 99 & 95 & 101 & 114 \\ 
    [1ex]
    & \multicolumn{24}{c}{$n = 1000$}\\
    \RomanI & 59 & 64 & 103 & 59 & 64 & 103 & 29 & 32 & 60 & 29 & 33 & 61 & 43 & 55 & 113 & 63 & 73 & 120 & 19 & 23 & 44 & 70 & 74 & 102 \\ 
    \RomanII & 73 & 76 & 129 & 73 & 76 & 128 & 64 & 65 & 109 & 64 & 65 & 109 & 61 & 65 & 114 & 64 & 68 & 118 & 75 & 78 & 123 & 67 & 70 & 117 \\ 
    \RomanIII & 88 & 90 & 126 & 88 & 90 & 125 & 65 & 67 & 96 & 64 & 66 & 95 & 68 & 72 & 108 & 82 & 91 & 130 & 131 & 132 & 158 & 94 & 95 & 127 \\ 
    \RomanIV & 68 & 71 & 83 & 67 & 71 & 83 & 35 & 38 & 46 & 35 & 38 & 46 & 41 & 52 & 81 & 53 & 59 & 79 & 34 & 38 & 46 & 74 & 78 & 89 \\   
    & \multicolumn{24}{c}{Scenarios with time-dependent covariates}\\
    & \multicolumn{24}{c}{$n = 200$}\\
    \RomanV & 77 & 84 & 104 & 77 & 84 & 105 & 67 & 71 & 86 & 67 & 71 & 86 & 151 & 140 & 138 & 152 & 139 & 135 & 71 & 75 & 91 & 167 & 149 & 126 \\ 
    \RomanVI & 100 & 107 & 167 & 100 & 106 & 168 & 67 & 69 & 123 & 66 & 69 & 123 & 166 & 153 & 170 & 150 & 152 & 191 & 122 & 183 & 304 & 222 & 218 & 251 \\ 
    \RomanVII & 116 & 119 & 179 & 116 & 119 & 179 & 91 & 94 & 146 & 91 & 94 & 146 & 150 & 144 & 176 & 160 & 167 & 215 & 131 & 165 & 242 & 190 & 188 & 221 \\ 
    [1ex]
    & \multicolumn{24}{c}{$n = 1000$}\\
    \RomanV & 64 & 65 & 74 & 64 & 65 & 75 & 57 & 57 & 69 & 56 & 57 & 69 & 152 & 137 & 133 & 155 & 138 & 134 & 59 & 59 & 64 & 158 & 139 & 122 \\ 
    \RomanVI & 87 & 91 & 146 & 85 & 89 & 145 & 55 & 57 & 108 & 55 & 56 & 108 & 187 & 174 & 190 & 157 & 159 & 200 & 62 & 89 & 230 & 215 & 211 & 234 \\ 
    \RomanVII & 108 & 112 & 165 & 108 & 115 & 170 & 84 & 85 & 132 & 83 & 84 & 132 & 162 & 153 & 179 & 162 & 174 & 227 & 92 & 103 & 162 & 182 & 174 & 202 \\ 
    \bottomrule
  \end{tabular}
\end{sidewaystable}

\vspace{-1cm}
\section{Application}
We illustrate the proposed methods through an application to a clinical trial conducted by Terry Beirn Community Programs for Clinical Research on AIDS \citep{abrams1994comparative,goldman1996response}. The trial was conducted to compare didanosine (ddI) and zalcitabine (ddC) treatments for HIV-infected patients who were intolerant to or had failed zidovudine treatments. Of the 467 patients recruited for the study, 230 were randomized to receive the ddI treatment, and the other 237 received the ddC treatment. The average follow up time was 15.6 months, and 188 patients died by the end of the study. Despite having longitudinal measurements that were measured at follow-up visits, \cite{abrams1994comparative} showed that the ddC treatment is more efficacious than the ddI treatment in prolonging survival time, based on a proportional hazards model with covariates measured at the baseline visit. We applied the proposed methods to investigate the time-dependent risk factors for overall survival. We included baseline covariates at randomization, such as gender, hemoglobin level, treatment received (ddI/ddC), and AIDS diagnosis (yes/no). We also included time-dependent covariates such as CD4 count, Karnofsky score, and cumulative recurrent opportunistic infections count. The CD4 count and Karnofsky score were measured at the baseline visit and bi-monthly follow-up visits. We adopted the last covariate carried forward approach between visit times when constructing these time-dependent covariates. For the opportunistic infection, we used the cumulative number of infections prior to $t$ as the covariate value at $t$. As in Remark 1, the Karnofsky score and CD4 count were transformed into the range $[0,1]$ using the corresponding empirical cumulative distribution functions.

\begin{figure}
	\centering
	\includegraphics{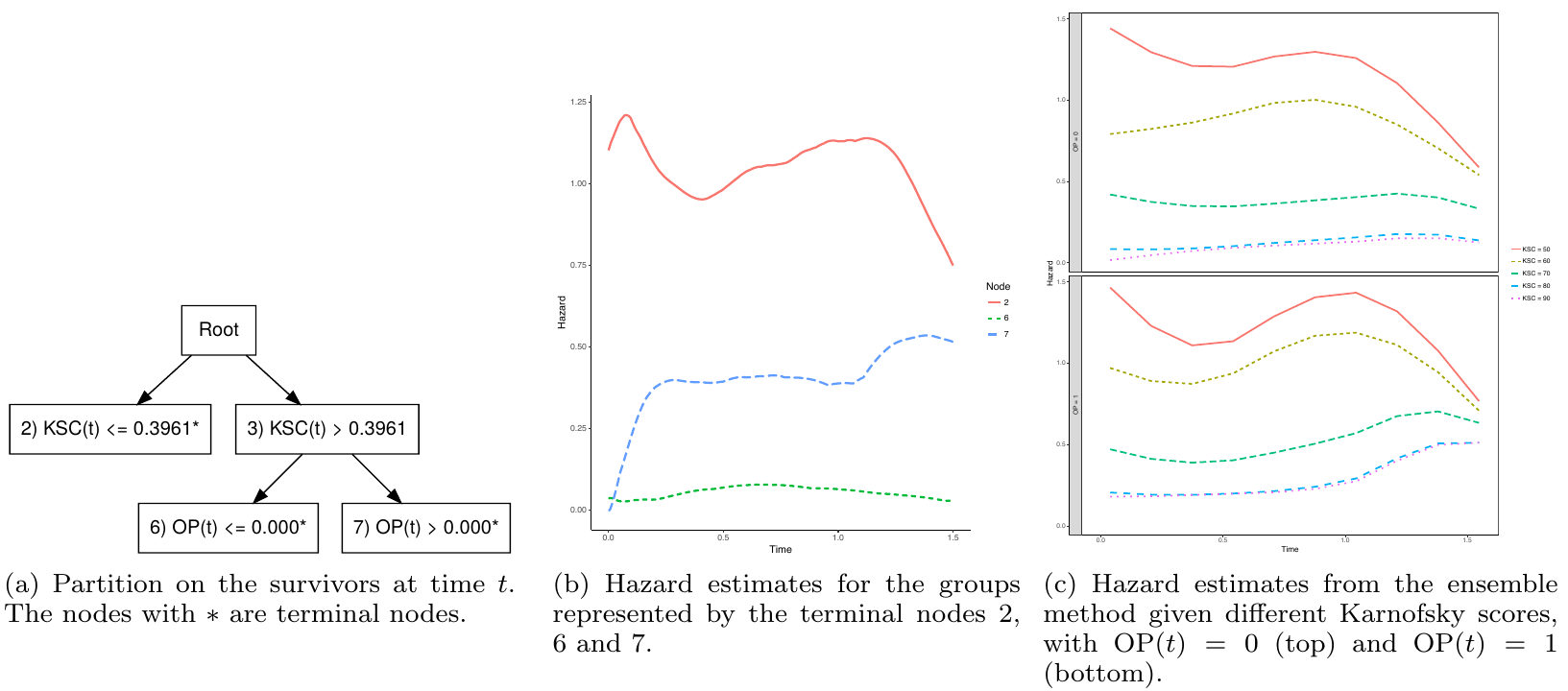}
	\caption{Mortality risks for the survivor population over time in the AIDS trial. This figure appears in color in the electronic version of this article, and any mention of color refers to that version.}
	\label{fig2}
\end{figure}

Figure~\ref{fig2}a displays the proposed ROC-guided survival tree. After pruning, the terminal nodes are $\tau_1 = \{\bZ(t) \mid \mbox{KSC}(t)\le 0.396\}$, $\tau_2 = \{\bZ(t)\mid \mbox{KSC}(t) > 0.396, \mbox{OP}(t) = 0\}$, and $\tau_3 = \{\bZ(t)\mid \mbox{KSC}(t) > 0.396, \mbox{OP}(t) > 0\}$, where KSC$(t)$ is the transformed Karnofsky score at $t$ and OP$(t)$ is the cumulative number of opportunistic infection up to $t$. Here the transformed Karnofsky score of $\mbox{KSC}(t) = 0.396$ corresponds to Karnofsky scores between 65--75 depending on $t$.
The partitions $\T = \{\tau_1, \tau_2, \tau_3\}$ corresponds to node 2, 6, and 7, whose estimated hazard rates are plotted in Figure~\ref{fig2}b. Figure~\ref{fig2}b clearly shows that a lower Karnofsky score is associated with higher mortality risk. For those with a high Karnofsky score, previous opportunistic infections are also associated with higher mortality risk. 

Although treatment received is not selected as a splitting variable, the effect of the treatment on survival was predominantly mediated through the Karnofsky score and opportunistic infections. Thus the survival tree provides insight on the mechanism by which the treatment operates. Moreover, the CD4 count is not used for splitting in the final tree, indicating that CD4 count may not be a surrogate endpoint in these patients; the result is consistent with the findings in \cite{goldman1996response}. For comparison, the result from \texttt{rpart} that holds time-dependent covariates at baseline is given in the Supporting Information. \texttt{rpart} also makes a split using the Karnofsky score at the root node but yields a larger tree.

Finally, we applied the proposed ensemble algorithm. To visualize the hazard estimation from the survival ensemble, we plot the hazard functions over time for different Karnofsky scores and two cumulative opportunistic infection counts while holding all other covariates constant at the median (or the mode for binary covariates). The hazard curves in Figure \ref{fig2}c show that a low Karnofsky score ($<70$ and dependent in normal activities) is associated with a higher risk of death, providing consistent results with our survival tree. On the other hand, for those with high Karnofsky scores ($>70$ and independent in normal activities), the hazard estimates are fairly flat but can increase after the occurrence of opportunistic infection episodes.

\section{Discussion}
In this article, we propose a unified framework for survival trees and ensembles, where $\ROC^*_t$ and related summary measures guide the tree-growing algorithm. The proposed tree-based hazard estimators involve kernel smoothing and could result in biased estimation when the terminal nodes contain a small number of observations. However, with the proposed pruning procedure, very large trees with insufficient node sizes are not likely to be selected as the final model. In practice, if the true model is complex and a very large tree is needed to capture the underlying truth, we recommend the use of ensemble methods.

In the presence of time-dependent covariates, the node membership for the same subject could change over time. In practice, we usually evaluate partitions on a finite set of time points. The computational cost of our algorithm mainly depends on the sample size, the number of predictors, and the number of time points on which we evaluate the partitions. It has been shown that the computational cost of the CART algorithm is $O(pn\log(pn))$ \citep{sani2018computational}. As is discussed in the Supporting Information, the computational cost for our algorithm that deals with time-dependent covariates is roughly $O(pqn\log(qn))$, where $q$ is the number of time points on which $\CON_t$ is evaluated. Simulation studies showed $q=20$ yields reasonably good performances when the sample size ranges from 100 to 1000.

\vspace{-.4in}
\section*{Acknowledgements}

This work was supported by the National Institutes of Health grants U19 AG033655, R01 CA193888, R01 HL122212, and the Health Resources and Services Administration grant, T0BHP29302. 
\vspace{-.4in}

\bibliographystyle{biom} 
\bibliography{BIOM2018393M_R2}

\begin{thebibliography}{}

\bibitem[\protect\citeauthoryear{Abrams, Goldman, Launer, Korvick, Neaton,
  Crane, Grodesky, Wakefield, Muth, Kornegay, et~al\mbox{.}}{Abrams
  et~al.}{1994}]{abrams1994comparative}
Abrams, D.~I., Goldman, A.~I., Launer, C., Korvick, J.~A., Neaton, J.~D.,
  Crane, L.~R., Grodesky, M., Wakefield, S., Muth, K., Kornegay, S., et~al.
  (1994).
\newblock A comparative trial of didanosine or zalcitabine after treatment with
  zidovudine in patients with human immunodeficiency virus infection.
\newblock {\em New England Journal of Medicine} {\bf 330,} 657--662.

\bibitem[\protect\citeauthoryear{Athey, Tibshirani, and Wager}{Athey
  et~al.}{2018}]{athey2018generalized}
Athey, S., Tibshirani, J., and Wager, S. (2018).
\newblock Generalized random forests.
\newblock {\em The Annals of Statistics} Forthcoming.

\bibitem[\protect\citeauthoryear{Bacchetti and Segal}{Bacchetti and
  Segal}{1995}]{bacchetti1995survival}
Bacchetti, P. and Segal, M.~R. (1995).
\newblock Survival trees with time-dependent covariates: application to
  estimating changes in the incubation period of {AIDS}.
\newblock {\em Lifetime Data Analysis} {\bf 1,} 35--47.

\bibitem[\protect\citeauthoryear{Bou-Hamad, Larocque, and Ben-Ameur}{Bou-Hamad
  et~al.}{2011a}]{bou2011discrete}
Bou-Hamad, I., Larocque, D., and Ben-Ameur, H. (2011a).
\newblock Discrete-time survival trees and forests with time-varying
  covariates: application to bankruptcy data.
\newblock {\em Statistical Modelling} {\bf 11,} 429--446.

\bibitem[\protect\citeauthoryear{Bou-Hamad, Larocque, and Ben-Ameur}{Bou-Hamad
  et~al.}{2011b}]{bou2011review}
Bou-Hamad, I., Larocque, D., and Ben-Ameur, H. (2011b).
\newblock A review of survival trees.
\newblock {\em Statistics Surveys} {\bf 5,} 44--71.

\bibitem[\protect\citeauthoryear{Breiman}{Breiman}{1996}]{breiman1996bagging}
Breiman, L. (1996).
\newblock Bagging predictors.
\newblock {\em Machine learning} {\bf 24,} 123--140.

\bibitem[\protect\citeauthoryear{Breiman}{Breiman}{2001}]{breiman2001random}
Breiman, L. (2001).
\newblock Random forests.
\newblock {\em Machine Learning} {\bf 45,} 5--32.

\bibitem[\protect\citeauthoryear{Breiman, Friedman, Stone, and Olshen}{Breiman
  et~al.}{1984}]{breiman1984classification}
Breiman, L., Friedman, J., Stone, C.~J., and Olshen, R.~A. (1984).
\newblock {\em Classification and Regression Trees}.
\newblock New York: Chapman \& Hall.

\bibitem[\protect\citeauthoryear{Chen, Hollander, and Langberg}{Chen
  et~al.}{1982}]{chen1982small}
Chen, Y., Hollander, M., and Langberg, N. (1982).
\newblock Small-sample results for the kaplan-meier estimator.
\newblock {\em Journal of the American statistical Association} {\bf 77,}
  141--144.

\bibitem[\protect\citeauthoryear{Ciampi, Thiffault, Nakache, and
  Asselain}{Ciampi et~al.}{1986}]{ciampi1986stratification}
Ciampi, A., Thiffault, J., Nakache, J.-P., and Asselain, B. (1986).
\newblock Stratification by stepwise regression, correspondence analysis and
  recursive partition: a comparison of three methods of analysis for survival
  data with covariates.
\newblock {\em Computational Statistics \& Data Analysis} {\bf 4,} 185--204.

\bibitem[\protect\citeauthoryear{Cui, Zhu, Zhou, and Kosorok}{Cui
  et~al.}{2019}]{cui2017some}
Cui, Y., Zhu, R., Zhou, M., and Kosorok, M. (2019+).
\newblock Consistency of survival tree and forest models: splitting bias and
  correction.
\newblock {\em arXiv preprint arXiv:1707.09631} .

\bibitem[\protect\citeauthoryear{Fisher and Lin}{Fisher and
  Lin}{1999}]{fisher1999time}
Fisher, L.~D. and Lin, D.~Y. (1999).
\newblock Time-dependent covariates in the {C}ox proportional-hazards
  regression model.
\newblock {\em Annual Review of Public Health} {\bf 20,} 145--157.

\bibitem[\protect\citeauthoryear{Fu and Simonoff}{Fu and
  Simonoff}{2017}]{fu2017survival}
Fu, W. and Simonoff, J.~S. (2017).
\newblock Survival trees for left-truncated and right-censored data, with
  application to time-varying covariate data.
\newblock {\em Biostatistics} {\bf 18,} 352--369.

\bibitem[\protect\citeauthoryear{Goldman, Carlin, Crane, Launer, Korvick,
  Deyton, and Abrams}{Goldman et~al.}{1996}]{goldman1996response}
Goldman, A.~I., Carlin, B.~P., Crane, L.~R., Launer, C., Korvick, J.~A.,
  Deyton, L., and Abrams, D.~I. (1996).
\newblock Response of {CD4} lymphocytes and clinical consequences of treatment
  using dd{I} or dd{C} in patients with advanced {HIV} infection.
\newblock {\em Journal of Acquired Immune Deficiency Syndromes} {\bf 11,}
  161--169.

\bibitem[\protect\citeauthoryear{Gordon and Olshen}{Gordon and
  Olshen}{1985}]{gordon1985tree}
Gordon, L. and Olshen, R.~A. (1985).
\newblock Tree-structured survival analysis.
\newblock {\em Cancer Treatment Reports} {\bf 69,} 1065--1069.

\bibitem[\protect\citeauthoryear{Harrell, Califf, Pryor, Lee, and
  Rosati}{Harrell et~al.}{1982}]{harrell1982evaluating}
Harrell, F.~E., Califf, R.~M., Pryor, D.~B., Lee, K.~L., and Rosati, R.~A.
  (1982).
\newblock Evaluating the yield of medical tests.
\newblock {\em Journal of the American Medical Association} {\bf 247,}
  2543--2546.

\bibitem[\protect\citeauthoryear{Heagerty and Zheng}{Heagerty and
  Zheng}{2005}]{heagerty2005survival}
Heagerty, P.~J. and Zheng, Y. (2005).
\newblock Survival model predictive accuracy and {ROC} curves.
\newblock {\em Biometrics} {\bf 61,} 92--105.

\bibitem[\protect\citeauthoryear{Hothorn, B{\"u}hlmann, Dudoit, Molinaro, and
  Van Der~Laan}{Hothorn et~al.}{2006}]{hothorn2006survival}
Hothorn, T., B{\"u}hlmann, P., Dudoit, S., Molinaro, A., and Van Der~Laan,
  M.~J. (2006).
\newblock Survival ensembles.
\newblock {\em Biostatistics} {\bf 7,} 355--373.

\bibitem[\protect\citeauthoryear{Hothorn, Lausen, Benner, and
  Radespiel-Tr{\"o}ger}{Hothorn et~al.}{2004}]{hothorn2004bagging}
Hothorn, T., Lausen, B., Benner, A., and Radespiel-Tr{\"o}ger, M. (2004).
\newblock Bagging survival trees.
\newblock {\em Statistics in Medicine} {\bf 23,} 77--91.

\bibitem[\protect\citeauthoryear{Huang, Chen, and Soong}{Huang
  et~al.}{1998}]{huang1998piecewise}
Huang, X., Chen, S., and Soong, S.-J. (1998).
\newblock Piecewise exponential survival trees with time-dependent covariates.
\newblock {\em Biometrics} {\bf 54,} 1420--1433.

\bibitem[\protect\citeauthoryear{Ishwaran and Kogalur}{Ishwaran and
  Kogalur}{2019}]{rfsrc}
Ishwaran, H. and Kogalur, U. (2019).
\newblock {\em Fast unified random forests for survival, regression, and
  classification (RF-SRC)}.
\newblock R package version 2.9.1.

\bibitem[\protect\citeauthoryear{Ishwaran and Kogalur}{Ishwaran and
  Kogalur}{2010}]{ishwaran2010consistency}
Ishwaran, H. and Kogalur, U.~B. (2010).
\newblock Consistency of random survival forests.
\newblock {\em Statistics \& Probability Letters} {\bf 80,} 1056--1064.

\bibitem[\protect\citeauthoryear{Ishwaran, Kogalur, Blackstone, and
  Lauer}{Ishwaran et~al.}{2008}]{ishwaran2008random}
Ishwaran, H., Kogalur, U.~B., Blackstone, E.~H., and Lauer, M.~S. (2008).
\newblock Random survival forests.
\newblock {\em The Annals of Applied Statistics} {\bf 2,} 841--860.

\bibitem[\protect\citeauthoryear{Kalbfleisch and Prentice}{Kalbfleisch and
  Prentice}{2011}]{kalbfleisch2011statistical}
Kalbfleisch, J.~D. and Prentice, R.~L. (2011).
\newblock {\em The Statistical Analysis of Failure Time Data}.
\newblock New York: John Wiley \& Sons.

\bibitem[\protect\citeauthoryear{LeBlanc and Crowley}{LeBlanc and
  Crowley}{1992}]{leblanc1992relative}
LeBlanc, M. and Crowley, J. (1992).
\newblock Relative risk trees for censored survival data.
\newblock {\em Biometrics} {\bf 48,} 411--425.

\bibitem[\protect\citeauthoryear{LeBlanc and Crowley}{LeBlanc and
  Crowley}{1993}]{leblanc1993survival}
LeBlanc, M. and Crowley, J. (1993).
\newblock Survival trees by goodness of split.
\newblock {\em Journal of the American Statistical Association} {\bf 88,}
  457--467.

\bibitem[\protect\citeauthoryear{Lin and Jeon}{Lin and
  Jeon}{2006}]{lin2006random}
Lin, Y. and Jeon, Y. (2006).
\newblock Random forests and adaptive nearest neighbors.
\newblock {\em Journal of the American Statistical Association} {\bf 101,}
  578--590.

\bibitem[\protect\citeauthoryear{McIntosh and Pepe}{McIntosh and
  Pepe}{2002}]{mcintosh2002combining}
McIntosh, M.~W. and Pepe, M.~S. (2002).
\newblock Combining several screening tests: {O}ptimality of the risk score.
\newblock {\em Biometrics} {\bf 58,} 657--664.

\bibitem[\protect\citeauthoryear{Meinshausen}{Meinshausen}{2006}]{meinshausen2006quantile}
Meinshausen, N. (2006).
\newblock Quantile regression forests.
\newblock {\em Journal of Machine Learning Research} {\bf 7,} 983--999.

\bibitem[\protect\citeauthoryear{Molinaro, Dudoit, and Van~der Laan}{Molinaro
  et~al.}{2004}]{molinaro2004tree}
Molinaro, A.~M., Dudoit, S., and Van~der Laan, M.~J. (2004).
\newblock Tree-based multivariate regression and density estimation with
  right-censored data.
\newblock {\em Journal of Multivariate Analysis} {\bf 90,} 154--177.

\bibitem[\protect\citeauthoryear{Moradian, Larocque, and Bellavance}{Moradian
  et~al.}{2017}]{moradian2017l_1}
Moradian, H., Larocque, D., and Bellavance, F. (2017).
\newblock $\mbox{L}_1$ splitting rules in survival forests.
\newblock {\em Lifetime Data Analysis} {\bf 23,} 671--691.

\bibitem[\protect\citeauthoryear{M{\"u}ller}{M{\"u}ller}{1991}]{muller1991smooth}
M{\"u}ller, H.-G. (1991).
\newblock Smooth optimum kernel estimators near endpoints.
\newblock {\em Biometrika} {\bf 78,} 521--530.

\bibitem[\protect\citeauthoryear{Muller and Wang}{Muller and
  Wang}{1994}]{muller1994hazard}
Muller, H.-G. and Wang, J.-L. (1994).
\newblock Hazard rate estimation under random censoring with varying kernels
  and bandwidths.
\newblock {\em Biometrics} {\bf 50,} 61--76.

\bibitem[\protect\citeauthoryear{Nobel}{Nobel}{1996}]{nobel1996histogram}
Nobel, A. (1996).
\newblock Histogram regression estimation using data-dependent partitions.
\newblock {\em The Annals of Statistics} {\bf 24,} 1084--1105.

\bibitem[\protect\citeauthoryear{Pe{\~n}a and Rohatgi}{Pe{\~n}a and
  Rohatgi}{1993}]{pena1993small}
Pe{\~n}a, E.~A. and Rohatgi, V.~K. (1993).
\newblock Small sample and efficiency results for the {N}elson-{A}alen
  estimator.
\newblock {\em Journal of Statistical Planning and Inference} {\bf 37,}
  193--202.

\bibitem[\protect\citeauthoryear{Pepe}{Pepe}{2003}]{pepe2003statistical}
Pepe, M.~S. (2003).
\newblock {\em The Statistical Evaluation of Medical Tests for Classification
  and Prediction}.
\newblock Oxford: Oxford University Press.

\bibitem[\protect\citeauthoryear{Sani, Lei, and Neagu}{Sani
  et~al.}{2018}]{sani2018computational}
Sani, H.~M., Lei, C., and Neagu, D. (2018).
\newblock Computational complexity analysis of decision tree algorithms.
\newblock In {\em International Conference on Innovative Techniques and
  Applications of Artificial Intelligence}, pages 191--197. Springer.

\bibitem[\protect\citeauthoryear{Schmid, Wright, and Ziegler}{Schmid
  et~al.}{2016}]{schmid2016use}
Schmid, M., Wright, M.~N., and Ziegler, A. (2016).
\newblock On the use of {H}arrell's {C} for clinical risk prediction via random
  survival forests.
\newblock {\em Expert Systems with Applications} {\bf 63,} 450--459.

\bibitem[\protect\citeauthoryear{Segal}{Segal}{1988}]{segal1988regression}
Segal, M.~R. (1988).
\newblock Regression trees for censored data.
\newblock {\em Biometrics} {\bf 44,} 35--47.

\bibitem[\protect\citeauthoryear{Steingrimsson, Diao, Molinaro, and
  Strawderman}{Steingrimsson et~al.}{2016}]{steingrimsson2016doubly}
Steingrimsson, J.~A., Diao, L., Molinaro, A.~M., and Strawderman, R.~L. (2016).
\newblock Doubly robust survival trees.
\newblock {\em Statistics in Medicine} {\bf 35,} 3595--3612.

\bibitem[\protect\citeauthoryear{Steingrimsson, Diao, and
  Strawderman}{Steingrimsson et~al.}{2019}]{steingrimsson2018censoring}
Steingrimsson, J.~A., Diao, L., and Strawderman, R.~L. (2019).
\newblock Censoring unbiased regression trees and ensembles.
\newblock {\em Journal of the American Statistical Association} {\bf 114,}
  370--383.

\bibitem[\protect\citeauthoryear{Therneau and Atkinson}{Therneau and
  Atkinson}{2018}]{rpart}
Therneau, T. and Atkinson, B. (2018).
\newblock {\em rpart: {R}ecursive partitioning and regression trees}.
\newblock R package version 4.1-13.

\bibitem[\protect\citeauthoryear{Therneau}{Therneau}{2015}]{survival-package}
Therneau, T.~M. (2015).
\newblock {\em A package for survival analysis in {S}}.
\newblock version 2.38.

\bibitem[\protect\citeauthoryear{Uno, Cai, Pencina, D'Agostino, and Wei}{Uno
  et~al.}{2011}]{uno2011c}
Uno, H., Cai, T., Pencina, M.~J., D'Agostino, R.~B., and Wei, L.~J. (2011).
\newblock On the {C}-statistics for evaluating overall adequacy of risk
  prediction procedures with censored survival data.
\newblock {\em Statistics in Medicine} {\bf 30,} 1105--1117.

\bibitem[\protect\citeauthoryear{Wager and Athey}{Wager and
  Athey}{2018}]{wager2017estimation}
Wager, S. and Athey, S. (2018).
\newblock Estimation and inference of heterogeneous treatment effects using
  random forests.
\newblock {\em Journal of the American Statistical Association} {\bf 113,}
  1228--1242.

\bibitem[\protect\citeauthoryear{Wallace}{Wallace}{2014}]{wallace2014time}
Wallace, M. (2014).
\newblock Time-dependent tree-structured survival analysis with unbiased
  variable selection through permutation tests.
\newblock {\em Statistics in Medicine} {\bf 33,} 4790--4804.

\bibitem[\protect\citeauthoryear{Zhang}{Zhang}{1995}]{zhang1995splitting}
Zhang, H. (1995).
\newblock Splitting criteria in survival trees.
\newblock {\em Statistical Modelling} {\bf 104,} 305--313.

\bibitem[\protect\citeauthoryear{Zhu and Kosorok}{Zhu and
  Kosorok}{2012}]{zhu2012recursively}
Zhu, R. and Kosorok, M.~R. (2012).
\newblock Recursively imputed survival trees.
\newblock {\em Journal of the American Statistical Association} {\bf 107,}
  331--340.

\end{thebibliography}

\section*{Supporting Information}
The mathematical details and proofs referenced in Sections 2,3 and 7, the additional simulation results referenced in Section 4, and the additional data analysis referenced in Section 5 are available with this paper at the Biometrics website on Wiley Online Library.

\end{document}